\documentclass[twocolumn,amsmath,amssymb]{revtex4}
\usepackage{graphicx}
\usepackage{dcolumn}
\usepackage{bm}
\begin{document}

\title{Incompressible States of Dirac Fermions in Graphene with 
Anisotropic Interactions}
\author{Vadim M. Apalkov}
\affiliation{Department of Physics and Astronomy, Georgia State University,
Atlanta, Georgia 30303, USA}
\author{Tapash Chakraborty$^\ddag$}
\affiliation{Department of Physics and Astronomy,
University of Manitoba, Winnipeg, Canada R3T 2N2}

\date{\today}
\begin{abstract}
We report on the properties of incompressible states of Dirac fermions in graphene 
in the presence of anisotropic interactions and a quantizing magnetic field. We 
introduce the necessary formalism to incorporate the anisotropy in the system.
The incopmpressible state in graphene is found to survive the anisotropy
upto a critical value of the anisotropy parameter. The anisotropy also introduces 
two branches in the collective excitations of the corresponding Laughlin state. 
It strongly influences the short-range behavior of the pair-correlation functions 
in the incompressible ground state.  
\end{abstract}
\maketitle

In his quest for a better understanding of the Laughlin state \cite{laughlin}, 
which is widely regarded as the best description of the fractional quantum Hall 
effect (FQHE) ground states \cite{FQHE_book} at the primary filling fractions 
($\nu=\frac1m$), Haldane \cite{haldane_11} recently demonstrated that the 
integer and the fractional quantum Hall effects are fundamentally different. In 
the latter case, he introduced a unimodular (area preserving) spatial metric that 
characterizes the shape of the correlation functions of the Laughlin state and 
is obtained by minimizing the correlation energy of the fractional quantum Hall state. 
This interaction metric is not necessarily congruent to the Galilean metric 
present in the one-body term of the system Hamiltonian. Such a geometric degree 
of freedom of the Hamiltonian is totally absent in the integer quantum Hall effect, 
but its presence helps to explain the success of the many-body state of Laughlin. 
Subsequent numerical studies \cite{haldane_12,haldane_2_12,fuchun} have elucidated 
various properties of the incompressible FQHE states in the presence of anisotropic 
interactions. Anisotropic transport in the FQHE regime is known to exist in higher 
order filling fractions \cite{aniso_expt}, and has also received some theoretical 
attention \cite{aniso_theory}. No such studies have been reported as yet for Dirac
fermions in graphene.

Of late, there has been an upsurge of interest on the magnetic field effects
on Dirac fermions in graphene \cite{abergeletal}. The FQHE states have
been investigated by us in monolayer \cite{monolayer_FQHE} and bilayer 
\cite{bilayer_FQHE} graphene. Experimentally, the presence of this effect in 
monolayer graphene has also been confirmed \cite{Abanin_10,Ghahari_11}. Interactions 
among Dirac fermions play a very important role in graphene, particularly in 
the quantum Hall regime \cite{ssc_Dirac}. Here we investigate the incompressible
state of graphene with anisotropic interactions. Our studies indicate that 
anisotropic interactions brings out several unique features in the system, in 
particular, in the collective modes and in the pair-correlation functions.

Monolayer graphene in a magnetic field $B$ has a discrete Landau level energy spectrum 
that is characterized by the Landau level index $n=0,\pm 1, \pm 2, \ldots $ and 
energy \cite{abergeletal,ssc_Dirac} $\varepsilon^{}_n=\hbar
\omega^{}_B \mbox{sgn}(n)\sqrt{|n|}$, where $\omega^{}_B=\sqrt2 v^{}_{\rm F}/\ell^{}_0$ 
and $\ell^{}_0 =\sqrt{\hbar/e B}$ is the magnetic length. Here $\mbox{sgn}(n) =0$ if 
$n=0$ and $\mbox{sgn}(n) =\pm 1$ if $n>0$ and $n<0$, respectively. The corresponding 
wave functions are of the form 
\begin{equation}
\Psi^{}_{n,m} =  C^{}_n
\left( \begin{array}{c}
 {\rm sgn}(n) {\rm i}^{|n|-1}\phi^{}_{|n|-1,m} \\
    {\rm i}^{|n|} \phi^{}_{|n|,m}   
\end{array}  
 \right),
\label{f1}
\end{equation}
where $C^{}_{n=0} = 1$ and  $C^{}_{n\neq 0} = 1/\sqrt2$. The functions $\phi^{}_{n,m}$ 
are the conventional Landau wavefunctions for a particle obeying the parabolic dispersion 
relation with the Landau index $n$ and $z$-component of electron angular momentum
$m$. Such conventional Landau wave functions are characterized by two sets of ladder 
operators: operator $b^{\dagger }$, which raises the Landau index $n$, and the guiding 
center ladder operator $a^{\dagger}$, which raises the intra-Landau index $m$. 
The energy spectrum depends only on $n$ and is highly degenerate with respect to the 
electron angular momentum $m$. In Eq.~(\ref{f1}) the Landau level 
wave functions are isotropic, i.e., the electron density depends only on 
$\rho = \sqrt{x^2 + y^2}$. The basis wave functions (\ref{f1}) are however, not unique. 
Due to degeneracy of the Landau levels, we can choose any single-particle basis, even 
anisotropic ones, to describe the properties of the electron system in a strong magnetic 
field. For the many-electron system with isotropic electron-electron interactions the 
wave functions (\ref{f1}) are the convenient basis for numerical analysis of the 
many-electron energy spectrum in a spherical geometry, since the isotropic potential 
conserve the angular momentum. In the spherical geometry \cite{haldane_83}, the interaction
properties of the many-electron system are described in terms of the Haldane 
pseudopotentials $V^{}_m$ \cite{haldane_83}, which are the energies of two electrons with 
relative angular momentum $m$. The radius of the sphere, $R$, is related to integer number
$2S$ of magnetic fluxes through the sphere in units of the flux quanta as $R = \sqrt{S} 
\ell^{}_0$. The single-electron states are characterized by the angular momentum, which 
is equal to $S$, and $S^{}_z=-S,\ldots, S$. For an isotropic potential, 
due to the spherical symmetry of the problem the many-particle states are described by the 
total angular momentum, $L$, and $L^{}_z$, while the energy depends only on $L$. 
As a result, we can evaluate the energy spectra of the system for a given value 
of $L^{}_z$, e.g., $L^{}_z = 0$ \cite{fano}, which greatly simplifies our analysis of 
the many-electron system in a given Landau level. 

\begin{figure}
\begin{center}\includegraphics[width=8cm]{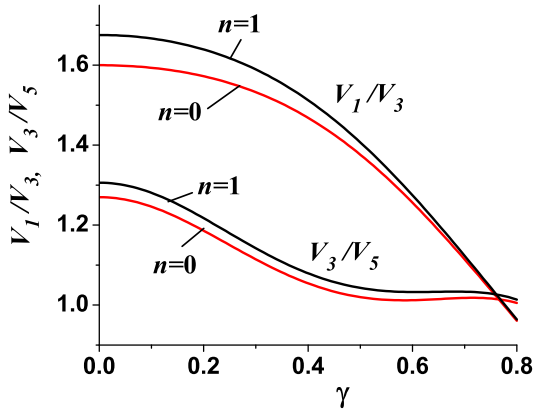}\end{center}
\vspace*{-1cm}
\caption{ The ratios of the Haldane's pseudopotentials, $V^{}_1/V^{}_3$ and
$V^{}_3/V^{}_5$ as a function of $\gamma$ in two graphene Landau levels
$n=0$ (red lines) and $n=1$ (black lines).
}
\label{Fig1}
\end{figure}

Here we consider the anisotropic electron-electron interaction potential of the type
\begin{equation}
V({\bf r}^{}_1, {\bf r}^{}_2) =\frac{e^2}{\kappa \sqrt{(x^{}_2-x^{}_1)^2 (1+\gamma) 
+ (y^{}_2-y^{}_1)^2 (1-\gamma)}},
\label{Van}
\end{equation}
where ${\bf r}^{}_1 = (x^{}_1,y^{}_1)$ and ${\bf r}^{}_2 =(x^{}_2,y^{}_2)$ are the
coordinates of two electrons, $\kappa$ is the dielectric constant, and 
the parameter $\gamma$ characterizes the anisotropy of the interaction potential with 
$\gamma = 0$ corresponding to isotropic interaction. We consider the properties of the 
many-electron system in graphene with fractional filling of the Landau level and with 
the anisotropic interaction (\ref{Van}). We assume that the interaction does 
not mix the states of different Landau levels. In this case the interaction potential  
should be projected on a given Landau level with index $n$. Although the interaction 
potential within a given Landau level is anisotropic, we can make it isotropic by
introducing a non-uniform scaling transformation of the coordinate system
\begin{equation}
x^{\prime } = x \sqrt{1+\gamma}; \ \ y^{\prime } = y \sqrt{1-\gamma}. 
\label{x1}
\end{equation}
The interaction then becomes isotropic $V \propto 1/|{\bf r}^{\prime}_1 - 
{\bf r}^{\prime }_2|^2$. To take advantage of isotropic potential in the scaled coordinate 
system, we need to define the angular momentum, which is conserved  by the isotropic 
interaction. Therefore we choose a single-electron basis within a 
given Landau level, which is initially anisotropic, and define the angular momentum in 
that anisotropic basis. This procedure was followed in Ref.\ \onlinecite{haldane_12}, where the anisotropic Landau 
level basis states with the non-Euclidean guide center metrics were introduced for 
the conventional Landau levels. These metrics are characterized by the anisotropy 
parameter $\gamma$. The corresponding guiding center intra-Landau level ladder operators 
are introduced through the Bogoliubov transformation  
$a^{}_{\gamma} = \frac1{\sqrt{1-\gamma^2}} \left(a + \gamma a^{\dagger}\right);
\ \ a_{\gamma }^{\dagger} = \frac{1}{\sqrt{1-\gamma^2}} \left(\gamma a + a^{\dagger} 
\right).$
The Landau level basis states of the non-relativistic system with parabolic dispersion
relation are then
\begin{equation}
\phi^{}_{n,m} (\gamma) = \frac{(b^{\dagger})^n (a^{\dagger}_{\gamma})^m}{\sqrt{n! m!}}  
\phi^{}_{0,0} (\gamma),
\label{psiG}
\end{equation}
where the state $\phi^{}_{0,0} (\gamma)$ is determined by the condition $a^{}_{\gamma} 
\phi^{}_{0,0} (\gamma )= b \phi^{}_{0,0} (\gamma) = 0$. For a given Landau level $n$ the 
basis function $\phi^{}_{n,m} (\gamma)$ is the linear combination of the isotropic basis 
functions $\phi^{}_{n,m} (\gamma=0)$. 

In the coordinate representation the zeroth conventional Landau-level function $\phi^{}_{0,0} (\gamma )$
of nonrelativistic electron is given by
\begin{equation}
\phi^{}_{0,0} (\gamma ) = \frac{(1-\gamma^2)^{1/4} }{\sqrt{2 \pi}} 
\exp \left( -\frac{1}{2} \gamma z^2 -\frac{1}{2} |z|^2 \right) ,
\end{equation}
where $z = x + {\rm i}y$ is the complex coordinate. Expressing the ladder operators in 
terms of the complex coordinate, $a = \tfrac12 z^{*} + \partial^{}_z$ and 
$b = \tfrac12 z + \partial^{}_{z^*}$, and using the expression for the anisotropic ladder 
operators $a^{}_{\gamma}$, we can construct the anisotropic Landau-level basis states from 
Eq.\ (\ref{psiG}). In graphene, the anisotropic Landau level basis functions are then  
\begin{equation}
\Psi^{}_{n,m} (\gamma ) =  C^{}_n
\left( \begin{array}{c}
 {\rm sgn}(n) {\rm i}^{|n|-1}\phi^{}_{|n|-1,m} (\gamma ) \\
    {\rm i}^{|n|} \phi^{}_{|n|,m} (\gamma )  
\end{array}  
 \right).
\label{f1G}
\end{equation}
The angular momentum in this anisotropic basis is $m$ and 
$L^{}_z (\gamma) = a^{\dagger} a$. For the many-electron system with anisotropic 
interaction and anisotropic basis states, the interaction properties are characterized 
by the Haldane pseudopotentials, $V^{}_m$, with relative angular momentum $m = m^{}_1 - m^{}_2$. Therefore the 
interaction energy of two electrons depends only on their relative momentum, but not 
on the total momentum, $M = m^{}_1 + m^{}_2$. With this property we can study the 
many-electron system with anisotropic interaction in a spherical geometry, where the 
total angular momentum is conserved.

We calculate $V^{}_m$ in a planar geometry for the wave functions in Eq.~(\ref{f1G}) 
and use these values in the spherical geometry to find the energy spectra of the 
many-electron system. It is easier to calculate the pseudopotentials $V^{}_m$ for
the two-electron system as the interaction energy of two electrons in a state with  
$M = 0$, and the relative momentum $m$: 
\begin{equation}
V^{}_m = \int d {\bf r}^{}_1  d {\bf r}^{}_2|\Phi^{}_m(\gamma, {\bf r}^{}_1, {\bf
r}^{}_2)|^2 V({\bf r}^{}_1, 
{\bf r}^{}_2),
\end{equation}
where the two-electron state $\Phi^{}_m(\gamma, {\bf r}^{}_1, {\bf r}^{}_2)$ with 
relative angular momentum $m$ is 
\begin{equation}
\Phi^{}_m(\gamma, {\bf r}^{}_1, {\bf r}^{}_2) = \frac{ (a_{\gamma, 1}^{\dagger}
-a_{\gamma, 2}^{\dagger})^m} {\sqrt{2^m m!}}\phi^{}_{0,0} (\gamma, {\bf r}^{}_1) 
\phi^{}_{0,0} (\gamma, {\bf r}^{}_2).
\end{equation}
Here $a_{\gamma, 1}^{\dagger}$ and $a_{\gamma, 2}^{\dagger}$ are the guiding center ladder 
operators for electrons 1 and 2, respectively. 
 
In graphene, the FQHE with a large many-particle excitation gap is for $n=0$ 
and $n=1$ Landau levels only, where the FQHE gap is the largest at $n=1$ Landau level
\cite{monolayer_FQHE,ssc_Dirac}. Here we consider only these two Landau levels.
To characterize the interaction properties of a partially occupied Landau 
levels, we study the FQHE at a filling factor $\nu = 1/3$. Similar behavior is 
expected for other filling factors, e.g., $\nu = 1/5$, $2/3$.

The wave function of the $n=0$ graphene Landau level is identical to the $n=0$ 
conventional Landau level of nonrelativistic system. Therefore in this case, the interaction properties and the 
pseudopotentials of graphene and conventional (non-relativistic) systems are identical.
The $n=1$ graphene Landau level is the mixture of $n=0$ and $n=1$ conventional Landau 
wave functions, which introduce specific features into the graphene system. 

\begin{figure}
\begin{center}\includegraphics[width=8cm]{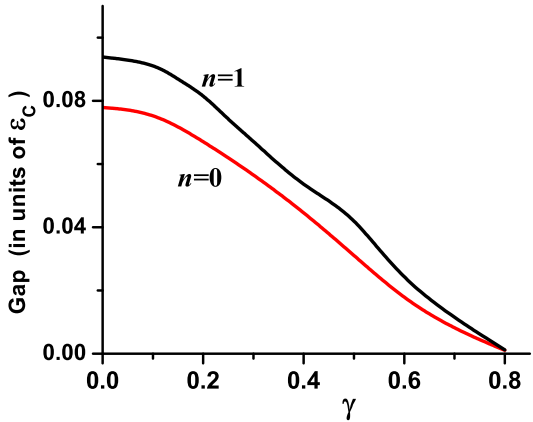}\end{center}
\vspace*{-1cm}
\caption{
The $\nu=\frac13$-FQHE gap as a function of $\gamma$ for the eight-electron system 
and $n=0$ (red line) and $n=1$ (black line). The FQHE gap is calculated in the spherical 
geometry with flux quanta $2S = 21$.
}
\label{Fig2}
\end{figure}
 
The magnitude of the FQHE gap depends on how $V^{}_m$ decreases with $m$. The 
FQHE with large gap is characterized by large values of the ratios $V^{}_1/V^{}_3$ 
and $V^{}_3/V^{}_5$. In Fig.\ \ref{Fig1} we show the ratio of the Haldane 
pseudopotentials as a function of $\gamma$. For all values of $\gamma$, the ratios 
of pseudopotentials are the largest at $n=1$ Landau 
level, which suggest that the FQHE state is more stable in $n=1$ Landau level. With 
increasing $\gamma$, the ratios of the pseudopotentials decrease 
for both $n=0$ and $n=1$ Landau levels, which makes the FQHE less stable at large 
$\gamma$. These results suggest that the FQHE gap decreases with increasing 
$\gamma$ and finally collapses for $\gamma \approx 0.8$, when the ratios of 
pseudopotentials are close to one.  
To characterize the dependence of the FQHE gap on anisotropy, 
we evaluate the energy spectra and excitation gap of a finite-size system consisting of $N=8$ 
electrons. The $\nu=\frac13$ FQHE gap is shown in Fig.\ \ref{Fig2} as a function of 
$\gamma$. With increasing $\gamma$ the FQHE gap decreases in both Landau levels, 
$n=0$ and $n=1$. The FQHE gap is larger for $n=1$. For $\gamma =\gamma^{}_{\rm cr} \approx 0.8$
the FQHE gap disappear and the $\nu = \frac13$ state becomes compressible. The critical value 
$\gamma^{}_{\rm cr}$ is the same for both Landau levels. This value corresponds 
to the condition that the pseudopotentials $V^{}_1$, $V^{}_3$, and $V^{}_5$ become 
almost the same (Fig.\ \ref{Fig1}). At small values of $\gamma < 0.15$ the FQHE gap shows weak dependence on 
anisotropy $\gamma$. 

In spherical geometry, which describes the isotropic system, the energy dispersion 
is obtained as a function of a total angular momentum, $L$. Each state has a $(2L+1)$ 
degeneracy. The typical energy spectrum is shown in Fig.\ \ref{Fig2}(a) for $\gamma 
= 0.4$. The energy spectrum has a finite gap and the low energy excited states of the 
spectrum has well defined single energy branch, showing a roton minimum at finite 
value of $L$  ($L=5$ in Fig.\ \ref{Fig2}(a)). In spherical geometry, the excited 
energy branch is described as a function of the angular momentum $E=E(L)$. Transition 
to the planar geometry is realized by replacing the angular momentum $L$ by the magnitude 
of the wave vector $k = L/R$, where $R$ is the radius of the sphere. In this case 
the energy of the lowest excited states depends on the magnitude of the wave vector 
but not on its direction, which corresponds to isotropic system. 

\begin{figure}
\begin{center}\includegraphics[width=8cm]{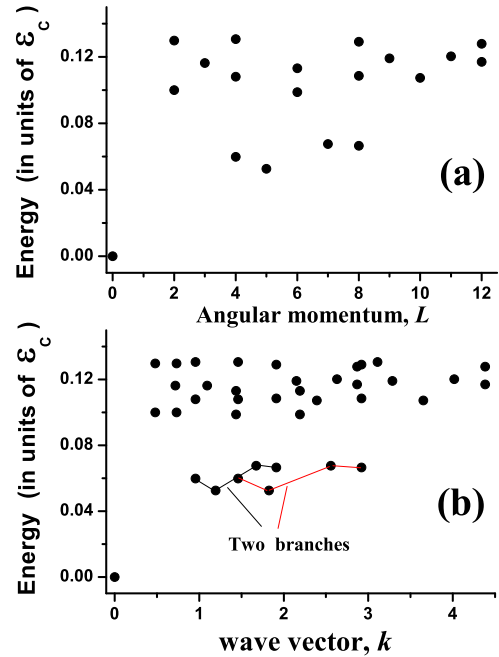}\end{center}
\vspace*{-1cm}
\caption{
(a) Energy spectrum of the eight-electron $\nu=\frac13$-FQHE system in the $n=1$
Landau level. The spectrum is calculated in the spherical geometry with flux quanta 
$2S = 21$, and $\gamma = 0.4$. (b) The energy spectrum of
the eight-electron $\nu=\frac13$-FQHE system for $n=1$ as a function of 
the wave vector, $k$ for $\gamma = 0.4$. The results are obtained 
from the energy spectrum in a spherical geometry and shown in panel (a). The two 
branches in (b) are shown schematically by black and red lines.
}
\label{Fig3}
\end{figure}

For anisotropic interactions with anisotropic basis, the energy depends not only on 
the magnitude of the wave vector $k$ but also on its direction. Such an anisotropic system can be made isotropic under a scaling 
coordinate transformation determined by Eq.\ (\ref{x1}). The corresponding 
transformation in the wave vector space is 
\begin{equation}
k_x^{\prime} = k^{}_x/ \sqrt{1+\gamma};
k_y^{\prime} = k^{}_y / \sqrt{1-\gamma}. \label{ky}
\end{equation}
Each state in the spherical geometry has $(2L+1)$ degeneracy. In the planar geometry 
these states correspond to $(2L+1)$ different directions of wave vector, 
$\beta ^{}_p = 2\pi p/(2L+1)$, where $p=1,\ldots, 2L+1$. The corresponding components of 
the wave vector are
$k_x^{\prime} = \frac LR \cos\phi^{}_n = \frac LR \cos\frac{2\pi p}{2L+1};
k_y^{\prime} = \frac LR \sin\phi^{}_n = \frac LR \sin\frac{2\pi p}{2L+1}.$
These are the components of the wave vector in the scaling coordinate system [Eqs.\ 
(\ref{ky})]. In this case the magnitude of the wave vector 
depends only on $L$: $(k_x^{\prime })^2 + (k_y^{\prime })^2 = (L/R)^2$ and does not 
depend on the direction. 

In the original coordinate system the components of the wave vector are, 
$k^{}_x = \sqrt{1+\gamma} k_x^{\prime} = \sqrt{1+\gamma} \frac LR \cos 
\frac{2\pi p}{2L+1};
k^{}_y = \sqrt{1-\gamma} k_y^{\prime} = \sqrt{1-\gamma} \frac LR \sin 
\frac{2\pi p}{2L+1},$
and the magnitude of the wave vector is 
$k^{}_p = \sqrt{k_x^2 +k_y^2} = \frac LR\sqrt{1 + \gamma\cos\frac{4\pi p}{2L+1}}.$ 
The magnitude of the wave vector now depends on the direction. In the original 
coordinate system, each energy level $E(L)$ with a given angular momentum $L$ produces 
$(2L+1)$ states with different wave vectors $k^{}_p$. In the thermodynamic limit 
($R\rightarrow \infty $), these wave vectors accumulate in two directions, corresponding 
to the points of large density, which is proportional to $1/(d k^{}_p/d p)$. This 
condition determines two values of the wave vector $k^{}_1 = \frac LR\sqrt{1 + \gamma}$ 
and $k^{}_2 = \frac{L}{R}\sqrt{1 - \gamma}$, which results in two branches in the 
low-energy dispersion relation of graphene. These branches are shown in Fig.\ \ref{Fig3}(b) 
for $\gamma = 0.4$. The energy dispersion shown in Fig.\ \ref{Fig3} (b) is recalculated 
from the energy spectrum obtained in the spherical geometry and shown in Fig.\ \ref{Fig3}(a). 
Splitting of the magneto-roton mode was first observed experimentally for conventional
semiconductor system \cite{roton_split} and was explained as a result of intrinsic 
anisotropy in the system \cite{tokatly_vignale}. Experimental confirmation of such
splitting in graphene would be very an important step forward.

\begin{figure}
\begin{center}\includegraphics[width=8cm]{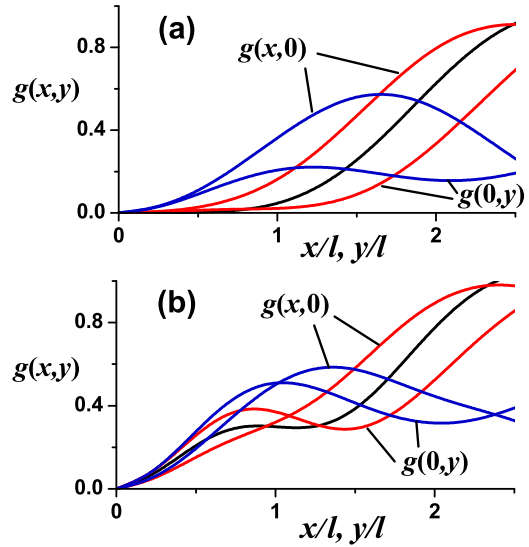}\end{center}
\vspace*{-1cm}
\caption{The pair-correlation function $g(x,y)$ for the eight-electron $\nu=\frac13$-FQHE 
system in the incompressible ground state for different anisotropy parameters:
$\gamma = 0$ (black line), $\gamma = 0.2$ (red line), and $\gamma = 0.6$ (blue line).
The results are for the spherical geometry with the flux quanta $2S = 21$.
We considered two Landau levels with indices (a) $n=0$ and (b) $n=1$. For 
$\gamma \neq 0$, the correlation function is anisotropic. For an anisotropic system, the
correlation functions are shown along the $x$ ($y=0$) and $y$ ($x=0$) directions.
}
\label{Fig4}
\end{figure}

Properties of the incompressible state can also be characterized by the pair-correlation 
function, 
\begin{equation}
g({\bf r} )=\int\ldots\int d {\bf r}^{}_3  \ldots d {\bf r}^{}_N |\Phi^{}_0 (\gamma,0, {\bf r}, 
{\bf r}^{}_3 , \ldots, {\bf r}^{}_N)|^2,
\end{equation}
where $\Phi^{}_0 (\gamma,{\bf r}^{}_1, {\bf r}^{}_2 ,{\bf r}^{}_3 , \ldots, {\bf r}^{}_N)$ 
is the $N$-particle wave function of the incompressible ground state. The ground state of 
the many-particle system is initially calculated in the spherical geometry for a special 
single-particle basis \cite{fano}, where the states of such basis are characterized by 
$L^{}_z$. Then each basis state of the spherical geometry, in the expression for the ground 
state of the many-electron system is replaced by the corresponding anisotropic state of the 
planar geometry with the same $z$ component of the angular momentum, $m=L_z$. The pair-correlation 
function in planar geometry is calculated in this way and are shown for different values
of $\gamma$ in Fig.\ \ref{Fig4}. 

For $\gamma = 0$, the pair-correlation function $g({\bf r})$ is 
isotropic and depends only on the magnitude of ${\bf r}$. For $n=0$
(Fig.\ \ref{Fig4}a), the incompressible state of the system is described by the Laughlin 
state, which results in $g(r) \propto r^6$ for small $r$. With increasing anisotropy, 
the correlation function becomes anisotropic and show quadratic dependence on $r$, 
which is correlated with suppression of the FQHE gap. Even for $\gamma =0.2$, when the 
FQHE gap is still large, the system shows a strong deviation of the pair-correlation 
function from the isotropic case. With increasing anisotropy the 
pair-correlation function shows local maxima for some finite values of $r$, which 
suggests a transition to compressible state with crystalline structure \cite{fuchun}.
For $n=1$ (Fig.\ \ref{Fig4}b), due to the presence of both $n=0$ and $n=1$ 
conventional Landau wave functions in a single-electron basis [Eq.\ (\ref{f1G})], 
the pair-correlation function has a quadratic dependence on $r$, even in the isotropic case.  
With increasing anisotropy, just as for $n=0$, the pair correlation 
function develops additional local maximum for finite values of $r$. 
The anisotropy of the pair-correlation function in $n=1$ is much weaker than
that for $n=0$, which suggests a more stable FQHE in the $n=1$ Landau level. This is 
consistent with the behavior of FQHE gap in different Landau levels [Fig.\ \ref{Fig3}].

The work has been supported by the Canada Research Chairs Program of the 
Government of Canada.

\end{document}